\documentclass[submission,copyright,creativecommons]{eptcs}

\providecommand{\event}{MARS 2020} 
\usepackage{breakurl}              
\usepackage{underscore}            

\usepackage{caption}
\usepackage[caption=false]{subfig}
\usepackage{tabularx, ltablex, booktabs, array}
\newcolumntype{C}{>{\centering\arraybackslash\hspace{0pt}}X}
\newcolumntype{P}[1]{>{\centering\arraybackslash\hspace{0pt}}p{#1}}

\usepackage{amsmath}
\usepackage{amssymb, wasysym}
\usepackage{enumitem}
\usepackage{subfig}
\usepackage{adjustbox}
\usepackage{verbatim}
\usepackage{lipsum}
\usepackage{multirow}
\usepackage{setspace}
\usepackage[table]{xcolor}
\usepackage{hhline}
\usepackage{algorithm}
\usepackage[noend]{algpseudocode}

\usepackage{verbatim}

\title{Synthesizing Strategies for Needle Steering in Gelatin Phantoms}

\author{
  Antje Rogalla${^{1\ast}}$ \qquad Sascha Lehmann${^{1\ast}}$ \qquad Maximilian Neidhardt${^{2}}$ \\ Johanna Sprenger${^{2}}$ \qquad Marcel Bengs${^{2}}$ \qquad Alexander Schlaefer${^{2}}$ \qquad Sibylle Schupp${^{1}}$
  \institute{${^1}$ Institute for Software Systems,\\ Hamburg University of Technology, Hamburg, Germany}
  \email{\{antje.rogalla, s.lehmann, schupp\}@tuhh.de}
  \institute{${^2}$ Institute of Medical Technology,\\ Hamburg University of Technology, Hamburg, Germany}
  \email{\{maximilian.neidhardt, johanna.sprenger, marcel.bengs, schlaefer\}@tuhh.de}\\[0.2cm]
  \textit{\small $^\ast$ Authors contributed equally}\\
  \small\textbf{Funding} This study was partially funded by the Technical University of Hamburg \\
  \small i$^3$~lab initiative (internal funding id T-LP-E01-WTM-1801-02).\\}
\def\titlerunning{Synthesizing Strategies for Needle Steering}
\def\authorrunning{A. Rogalla \& S. Lehmann}
\begin{document}
\maketitle

\begin{abstract}
In medicine, needles are frequently used to deliver treatments to subsurface targets or to take tissue samples from the inside of an organ.
Current clinical practice is to insert needles under image guidance or haptic feedback, although that may involve reinsertions and adjustments since the needle and its interaction with the tissue during insertion cannot be completely controlled.
(Automated) needle steering could in theory improve the accuracy with which
a target is reached and thus reduce surgical traumata especially for minimally invasive procedures, e.g., brachytherapy or biopsy.
Yet, flexible needles and needle-tissue interaction are both complex and expensive to model and can often be computed approximatively only.
In this paper we propose to employ timed games to navigate flexible needles with a bevel tip to reach a fixed target in tissue.
We use a simple non-holonomic model of needle-tissue interaction, which abstracts in particular from the various physical forces involved and appears to be simplistic compared to related models from medical robotics.
Based on the model, we synthesize strategies from which we can derive sufficiently precise motion plans to steer the needle in soft tissue.
However, applying those strategies in practice, one is faced with the problem of an unpredictable behavior of the needle at the initial insertion point.
Our proposal is to implement a preprocessing step to initialize the model based on data from the real system, once the needle is inserted.
Taking into account the actual needle tip angle and position, we generate strategies to reach the desired target.
We have implemented the model in Uppaal Stratego and evaluated it on steering a flexible needle in gelatin phantoms;
gelatin phantoms are commonly used in medical technology to simulate the behavior of soft tissue.
The experiments show that strategies can be synthesized for both generated and measured needle motions with a maximum deviation of $1.84mm$.
\end{abstract}

\section{Introduction}

In many medical procedures needle interventions are indispensable.
Needles are used for minimally invasive medical procedures, including brachytherapy in radiation therapy, as well as for diagnostic procedures as biopsy.
Due to their small diameter, needles can reach subsurface targets within the organ.
Needles that are flexible have the additional ability of steering around sensitive tissues or anatomic obstacles.
Using minimal surgical techniques, a flexible needle can thus reach the target inside soft tissue with little trauma and is therefore  used to deliver drugs or radioactive seeds, or for thermal ablation.
Obviously, an important property in most medical applications is the accuracy with which the needle tip reaches a target
since otherwise results are false negatives (e.g., in of case biopsy) or flatly unsound (e.g., in case of radioactive seeds implementing during brachytherapy, when the healthy tissue is targeted instead of cancerous tissue).
However, the current practice in clinical settings is manual needle placement under image guidance such as ultrasound or even by haptic feedback.
Since manual insertion has only limited control of the needle-tissue interaction, both the resulting accuracy of needle-tip placement varies and soft tissue injuries increase through reinsertions and adjustments of the needle.

\medskip
In recent years, several methods of needle steering have been proposed to control algorithmically the motion of needles inside of tissue.
Since the ultimate positioning of a needle tip depends not only on the insertion location but also on needle bending and tissue deformation, research on steerable needles encompasses technologies for device design as well as methods for modeling, path planning, or control procedures.
The major challenge in needle steering is in the uncertainty of needle motion due to tissue deformation, inhomogeneities, and a number of other control and environment parameters during needle intervention.
As a consequence, flexible needles and needle-tissue interaction are complex and expensive to model and current work on motion planning and control for tip-steerable needles employs expensive numerical methods to compute optimal motion paths.

In this paper we propose to use timed games for navigating flexible needles with a bevel tip to reach a fixed target in tissue.
In that game, the flexible needle with bevel tip is controllable and acts as the protagonist. The needle can move forward or rotate and, in case of rotation, changes its direction of movement.
Soft tissue---in our application represented by gelatin phantoms---is non-controllable and acts as the antagonist in the game.
The winning condition of the game is to reach a target in tissue with fixed accuracy.
The protagonist, the needle, wins the game, if a strategy for the needle exists so that the needle always reaches the target regardless of how the tissue behaves.

We have implemented needle-tissue interaction as timed game model in Uppaal Stratego, which synthesizes strategies from which motion plans for the needle can be derived.
In this paper, we report on the Uppaal model and the strategies as well as on the issues we encountered when applying the strategy in practice.
Particularly, the unpredictable behavior of the needle at the initial insertion point presented a problem to us.
We propose a pre-processing step that initializes the model based on data from the running system, after the needle has been inserted, and use the actual needle tip angle and position as input for generating strategies that reach the desired target.
Based on the synthesized strategies, we provide needle motion plans in soft tissue.
Our experiments show that needle motion planning by strategy synthesis is sufficiently precise.
For both generated and measured needle motion traces, strategies could be synthesized with a maximum deviation of $1.84mm$.\\

The outline of the paper is as follows. First, we provide an overview of related work (Sec.~\ref{sec:related-work}).
In Sec.~\ref{sec:medical-background} we describe the medical background and the experimental setting.
Then, we introduce the system model (Sec.~\ref{sec:system-and-model}), followed by a description of strategy synthesis on that model (Sec.~\ref{sec:strategy-synthesis}).
We conduct two types of experiments on generated and observed data in Sec.~\ref{sec:experiments}, and discuss possible threats to validity.
Finally, we conclude  in Sec.~\ref{sec:conclusion}.

\section{Related Work} \label{sec:related-work}

In medical robotics, needle steering is a subfield of its own.
There, one typically investigates application-specific, or at least highly specialized, configurations of needle tip, needle material, and tissue.
A study in Nature \cite{Van:2017}, for example, showed how different needle tip designs influence tip-tissue contact forces.
Other work shows various types of motion planning and control algorithms for needle steering.
Fu et al. \cite{Fu:2018} use CT-imaging and extract cost maps for planning needle paths that avoid certain anatomical structures for lung biopsies.
Mathematically speaking, the models are, among others, based on optimization \cite{Duindam:2010, Sun:2016}, fractal trees \cite{Liu:2016, Pinzi:2019}, or sampling \cite{Sun:2015}.
An overview of the technological and algorithmic state-of-the-art in robotic needle steering including various aspects of modeling needle-tissue interaction, needle paths, and motion planning is presented in  Cowan et al.~ \cite{Cowan:2011}.
This overview also discusses non-holonomic, stochastic, torsional as well as mechanics-based modeling approaches.
Following \cite{Webster:2006}, we utilize the non-holonomic modeling of the needle-tissue interaction.

As formal notation, we choose timed games.
Timed games combine classical model checking \cite{Alur:1994} and controller synthesis \cite{Cassez:2005}, and do not only offer the possibility to prove system properties, but also to generate strategies so that desired system properties are fulfilled.
A timed game consists of a timed game automata system, where the edges are partitioned into controllable and uncontrollable, and a winning condition, e.g., a safety or reachability property.
The associated control synthesis problem of a timed game asks if the timed game automaton is controllable in a way so that the winning condition is always satisfied.
Timed games provide a suitable framework for the analysis of real-time systems \cite{Faella:2014}.
Established tools for automatic control synthesis of real-time systems modeled as timed game automata include Uppaal Tiga \cite{Behrmann:2007} and Uppaal Stratego \cite{David:2015}.
Timed games using Uppaal Tiga have been applied to control climate control systems, oil pumps, and data paths of printers and copiers \cite{Jessen:2007}, \cite{Cassez:2009}, \cite{AlAttili:2009};
Uppaal Stratego is successfully applied, for example,  for the synthesis of near-optimal traffic light controllers \cite{Eriksen:2017}. In the field of medical technology, timed games have not yet been applied.

\section{Medical background} \label{sec:medical-background}

\subsection{Needle Steering}
Placing needles precisely at a target location inside a soft tissue is a daily clinical task.
Commonly, the physician relies on ultrasound imaging during needle placement.
Thereby, target and needle tip are focused in the image while the needle penetrates the tissue.
The physician needs to avoid fragile structures such as nerves during insertion.
Hence, a direct path between needle tip and target location is many times not feasible.
Consequently, the physician needs to correct insertions by retracting and re-orienting the needle shaft leading to lengthy procedures.
Needle placing is even more challenging during local anesthetic nerve blocks where the physician needs to rinse the volume around a nerve with anesthetics while avoiding injuries to the nerve itself and reaching regions located behind the nerve.

We propose a flexible needle design with a $45^{\circ}$ beveled needle tip, which allows us to steer the needle inside the tissue avoiding fragile structures and reaching target regions without retractions.
However, steering a needle inside tissue is challenging since ultrasound only offers two dimensional information of a three dimensional setup.
Hence, strategy synthesis can assist surgeons in finding designated paths.

\begin{figure*}[t!]
   \centering
   \begin{minipage}{.33\textwidth}
       \centering
       \includegraphics[width=.90\linewidth]{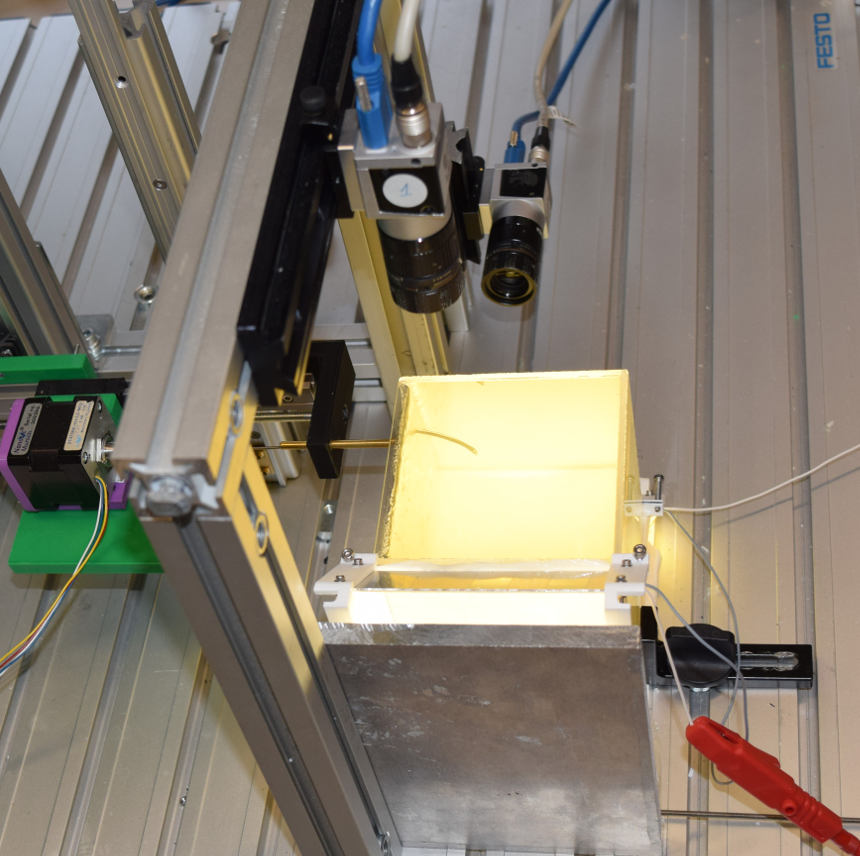}
   \end{minipage}
   \begin{minipage}{.47\textwidth}
       \centering
       \includegraphics[width=.90\linewidth]{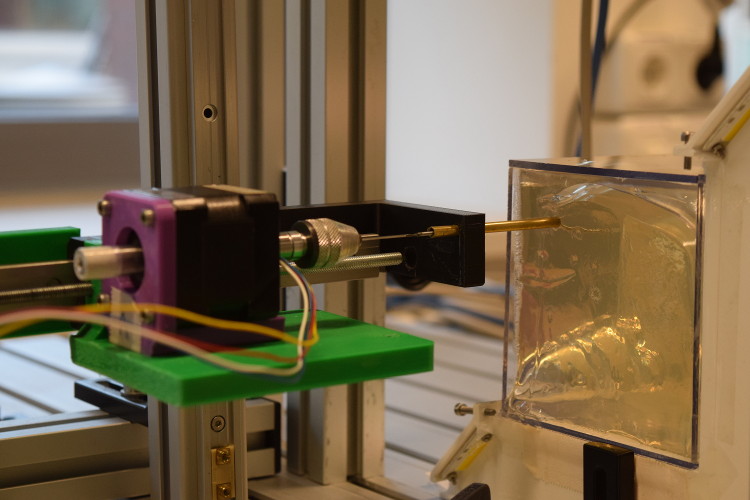}
   \end{minipage}%
\caption{Experimental Setup.
Left: Two cameras track the needle tip during insertions into a gelatin block.
Right: A deformable needle with a bevel tip is guided through hollow shaft stepper motor and hence can be rotated around its axis for steering.
A second stepper motor (not visible in the image) forwards the needle into the phantom.}
\label{fig:experimental-setup}
\end{figure*}

\begin{figure*}[b!]
   \centering
   \begin{minipage}{.30\textwidth}
       \centering
       \includegraphics[width=.90\linewidth]{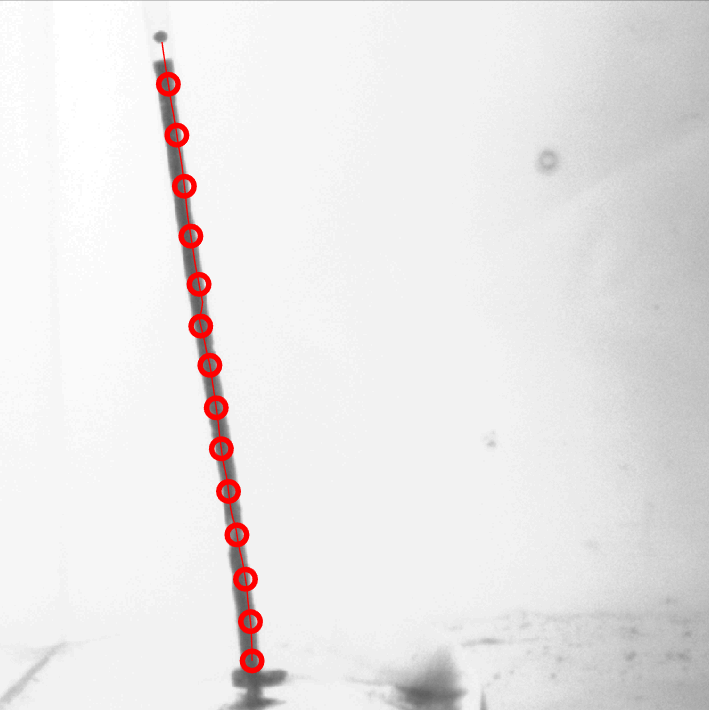}
   \end{minipage}
   \begin{minipage}{.30\textwidth}
       \centering
       \includegraphics[width=.90\linewidth]{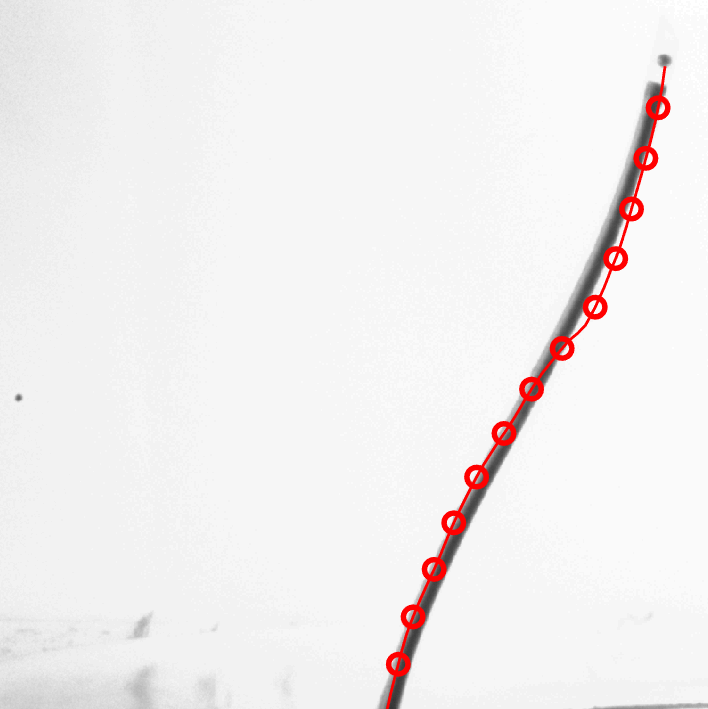}
   \end{minipage}%
   \caption{Example images of the cameras. The red circles indicate a selection of previously detected positions of the needle tip. The needle tip can be identified by the steel ball marker. In this example the needle is rotated once resulting in an s-shaped path.}
\label{fig:needle-images}
\end{figure*}

\subsection{Experimental Data Acquisition}
We evaluate the accuracy of our model with an experimental setup for inserting needles inside tissue mimicking phantoms, shown in Figure~\ref{fig:experimental-setup}.
We designed a custom resin needle tip with a $45^{\circ}$ inclination.
A steel ball was molded at the center for tracking purposes.
The tip was attached to a flexible shaft, which allows us to change the orientation of the tip inside the tissue during insertions.
The needle was fixed to a hollow shaft stepper motor, which allows us to rotate the needle in $1.8^{\circ}$ steps.
This motor was attached to a stepper stage for needle insertion.
During all experiments the needle was inserted with a velocity of $0.7~mm/sec$.
We use gelatin as a tissue mimicking phantom with a $10\%$ gelatin to water concentration.
For tracking the needle tip we fixed two cameras (Basler acA1300-200uc) at an angle of approximately $90^{\circ}$ to each other.
The relative orientation of the two cameras was estimated by a checkerboard calibration in water \cite{Bradski:2000}.
To compute the three dimensional position of our needle tip, we first use image processing methods to detect the pixel coordinates of the steel ball (see Figure~\ref{fig:needle-images}), which is located at the needle tip center.
Next, the coordinates were undistorted and the three dimensional position of the steel ball was estimated by mid-point-triangulation.
A digital oscilloscope (Saleae, Logic Analyzer) records imaging and motor step triggers during all experiments.
Hence, we can map the estimated three dimensional positions to the designated motor steps.

\section{System Description and Model} \label{sec:system-and-model}

In this section, we decide on a suitable underlying model for needle steering based on the requirements of the needle insertion device, and introduce an Uppaal model that allows checking the needle motion and synthesizes strategies for its navigation.
Finally, we will cover those aspects of our model that require additional consideration when implementing it in Uppaal.

\subsection{Needle Steering Model Choice}
For the choice of a suitable needle steering model for our motion planning task, several model types such as \textit{non-holonomic}, \textit{stochastic}, \textit{torsional}, and \textit{mechanics-based} \cite{Cowan:2011} models may be considered.
While the latter, physical models allow a more realistic representation of the actual phenomena of such systems, models based on geometric properties are often easier to implement, simulate, and verify.
Based on the described needle insertion device, the following requirements are relevant:
First, as the gelatin we use as substitution for real tissue is generally homogeneous and independent of the concrete insertion angle, we can omit a modeling of concrete tissue properties, and it is sufficient to consider the 2-dimensional projection of the needle motion.
Second, we restrict our model to two rotations at most, to reduce injuries caused by repeated needle adjustments.
Third, the velocity of the needle remains constant throughout the experiments, and the needle currently only allows forward motions.
Taking these requirements and assumptions into account, we decided on a non-holonomic model to reduce the process to the sole needle motion path, without considering additional influences of the needle-tissue interaction.
In the literature, it was shown that these simplified models can already match the needle motion (at least locally) sufficiently.
One such model---which we will use as a base of our implementation---is the \textit{bicycle model} \cite{Webster:2006}, which represents the needle as a two-wheel system, where the angle of the front wheel---representing the bevel tip---determines the concrete circle path that the needle tip follows.
In this model, only the geometry and the motion curve are considered, while the influence of external and internal forces, elasticity, and friction are neglected.

\begin{figure}[t]%
  \centering
  \includegraphics[width=1.0\linewidth]{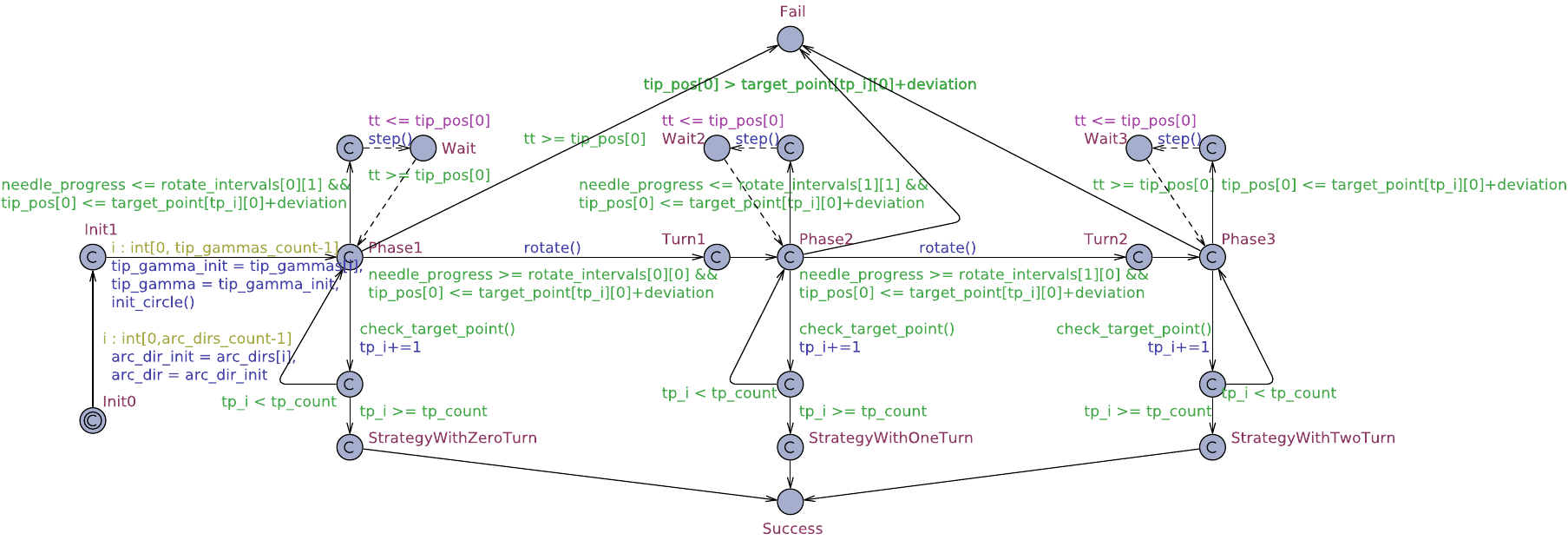}
	\caption{The main \texttt{Model} template of the Uppaal needle steering model}
	\label{fig:needle-steering-uppaal-model}
\end{figure}

\subsection{Uppaal Stratego Model}
For this paper, we use \textit{version 4.1.20-5} of \textit{Uppaal Stratego}.
The Uppaal implementation of the needle steering process is divided into a \textit{declaration}, the \textit{model templates}, and the \textit{queries} that should be checked on the model.
The \textit{declaration} part consists of the functions for the circle path initialization (\texttt{init_circle()}), the needle motion steps (\texttt{step()}), and the needle rotation (\texttt{rotate()}).
Furthermore, it contains arrays that hold the potential initial needle tip angles, circle arc directions and the real observation data, as well as the concrete target points which the needle should cross with the navigation plans we want to derive from the model.
In addition, it contains all adaptable parameters such as the rotation time intervals, amounts and deviations of data points, the x and y coordinates of the needle tip position (\texttt{tip_pos[0], tip_pos[1]}), angle, velocity, and the time step size.

The \textit{model templates} part consists of the main \texttt{Model} and a helper \texttt{Trace} template.
The \texttt{Model} (abbr. M) shown in Figure~\ref{fig:needle-steering-uppaal-model} is divided into initialization edges, on which an arc direction and an insertion angle are selected from the given arrays via \texttt{select} statements, and the core motion functionality, which consists of the following locations:
\begin{itemize}
  \item \texttt{PhaseX}: A choice between rotation and no rotation for the current time step is made (controllable). If no rotation is chosen, an uncontrollable \texttt{step} is executed
  \item \texttt{Wait}: Waiting for a specific amount of time until the next step / rotation decision is made
  \item \texttt{TurnX}: Added to support the identification of rotation points in the generated strategy data (no impact on the model functionality)
  \item \texttt{Fail}: Activated if the target region cannot be reached anymore
  \item \texttt{StrategyWithXTurn}: Helper locations which allow defining strategies over specific rotations
  \item \texttt{Success}: Activated if the target region defined by the last target point is reached (otherwise - if more target points exist - the next point is read and \texttt{PhaseX} becomes active again)
\end{itemize}
To avoid destroying the tissue, we do not allow more than two rotations in our model, and model the movement sequences of the needle after each rotation via individual locations.
Alternatively, the single phases could be merged to one by adding a counter.
We should note that not all phases need to be traversed, but only those whose rotations are relevant to reach the (set of) specified target points.
We underline that our model generally allows different initial angles and directions of the needle, but in our work the model is initialized with the real initial angle and direction.
The \texttt{Trace} template is an optional helper template, which steps through the observation trace, enabling us to plot real observed needle positions using the simulation functionalities of UPPAAL.

At last, the \textit{queries} allow us to check reachability for the existence and concrete instances of suitable navigation plans, and simulate arbitrary paths through the model.
We use the following three types of queries:
\begin{itemize}
  \item \texttt{E<> M.StrategyWithXTurn} and \texttt{A<> M.StrategyWithXTurn}:\\
        Symbolic model checking to determine if at least one path may lead to the target
  \item \texttt{simulate N [<= T] \{tip_pos[0], tip_pos[1]\}}:\\
        Test simulation runs to initially check possible traces generated by the model
  \item \texttt{strategy reachStrategyWithXTurn = control: A<> (M.StrategyWithXTurn)} and \\
        \texttt{simulate N [<= T] \{tip_pos[0], tip_pos[1]\} under reachStrategyWithXTurn}:\\
        Queries for the generation and simulation of strategies, further covered in Sec.~\ref{sec:strategy-synthesis}
\end{itemize}
For the model implementation in Uppaal, we have to consider additional aspects to allow for an application of the model to our needle motion task.
As the strategy synthesis and symbolic model checking queries do not support \texttt{double} variables, we have to convert all affected variables - at least for storing - into the \texttt{integer} domain.
Generally, we achieve this by scaling all double variables data by a certain magnitude (e.g., transforming $0.01$ into $10$ via the multiplication factor $1000$) and use the Uppaal function \texttt{fint()} to cast the scaled values to integers.
We transform them back into the double domain only for concrete calculations within of Uppaal.
Furthermore, the needle motion is discretized in Uppaal, so that the \texttt{step()} and \texttt{rotation()} functions are only executed every time delta $dt$.
Calling the step function, the needle feed and the $x$ and $y$ coordinate of the needle are updated depending on the choice of $dt$.
In our model one step corresponds to a needle feed of $0.7~mm/sec$.
Finally, as described before, a few helper locations are added to the model to allow for an easier generation and interpretation of strategies as well as a simplified conversion of strategies into needle motion plans.

\section{Strategy Synthesis} \label{sec:strategy-synthesis}

In needle steering, one lacks on the one hand information about the real system, e.g., soft tissue properties and the insertion angle or the precise characteristics of the needle. On the other hand, one has to generate a motion plan to navigate the needle to a specified target point with an error as small as possible. In this section, we present needle steering as a timed game and discuss the strategies we defined.

\subsection{The Timed Game System}
The system model of needle-tissue interaction described in detail in the previous chapter can be interpreted as a timed game automata system.
In that model, the controllable and uncontrollable edges (plain and dashed in Figure \ref{fig:needle-steering-uppaal-model}) define the interaction of the needle with its environment, the tissue.
As we can manipulate the needle rotation and movement, the transitions caused by the needle define the controllable edges of the model.
The uncontrollable edges of the model, on the other hand, are the transitions caused by the anatomic characteristics of tissue.
In the current model, the tissue is assumed to be homogeneous. Uncontrollable transitions therefore do not deflect the needle.
As soon as inhomogeneities of tissue or anatomic obstacles are taken into account, however, uncontrollable transitions will influence the needle motion; we discuss extensions to our current model in future work. \\

The winning condition in the timed game is that a target point in tissue is reached.
Accordingly, strategy synthesis seeks to generate motion sequences for the needle such that the winning condition is satisfied.
Such winning strategy does not always exist, because not all points in tissue can be reached by a bevel tip needle.
Since most needle-steering scenarios require rotating the needle, we could further constrain the winning condition, for example, by
determining the minimal distance between two rotation points, and we could devise cost-optimal or time-optimal winning strategies, for example, for minimizing the number of rotations, keeping the tissue intact.
Yet, we relegate those optimizations to future work.
In this paper, we focus on generating strategies with the smallest possible deviation to real target points.


\subsection{Synthesizing Strategies}
For our model, we formulate the winning condition (of reaching the final target point \texttt{target[n][0], \texttt{target[n][1]}} in tissue) within a window of half of the needle velocity defined via the deviation $dev$.

\begin{verbatim}
   (tip_pos[0]>=(target[n][0]-dev)) && (tip_pos[0]<=(target[n][0]+dev)) &&
   (tip_pos[1]>=(target[n][1]-dev)) && (tip_pos[1]<=(target[n][1]+dev))
\end{verbatim}

\noindent Assuming homogeneous tissue, a winning strategy exists whenever a geometric path exists between the insertion point and the target point with maximal two rotations of the needle.
With this formulation of the winning condition, however, we encountered practical problems in Uppaal.

Therefore, we formulate the property in the function \texttt{check_target_point() and reformulate the winning condition} including the locations \texttt{M.StrategyWithXTurn} from the model \texttt{M}, where \texttt{X} represents zero, one, or two rotations; note that the location is visited only if the target is reached by the needle with \texttt{X} rotations.
The query for generating the desired strategy then reads as:
\begin{center}
\texttt{strategy reachStrategyWithXTurn = control: A<> (M.StrategyWithXTurn)}
\end{center}

Assuming, again, homogeneous tissue, the winning strategy in its entirety corresponds to all motion plans that reach the target.
The rotation points of a specific motion plan in that set can be read off from the output of the strategy, which determines how far the needle should be fed without rotation and when the needle has to be rotated.
To reduce the number of motion plans, we restrict the insertion angle or the permitted deviation.
Desirable is a strategy by which the needle reaches the target as closely as possible, that is, where the permitted deviation is as small as possible.

\begin{figure}[b]%
  \centering
  \subfloat[All motion plans of the complete strategy]{
    \label{fig:entire-strategy-example}
    \includegraphics[width=0.45\linewidth]{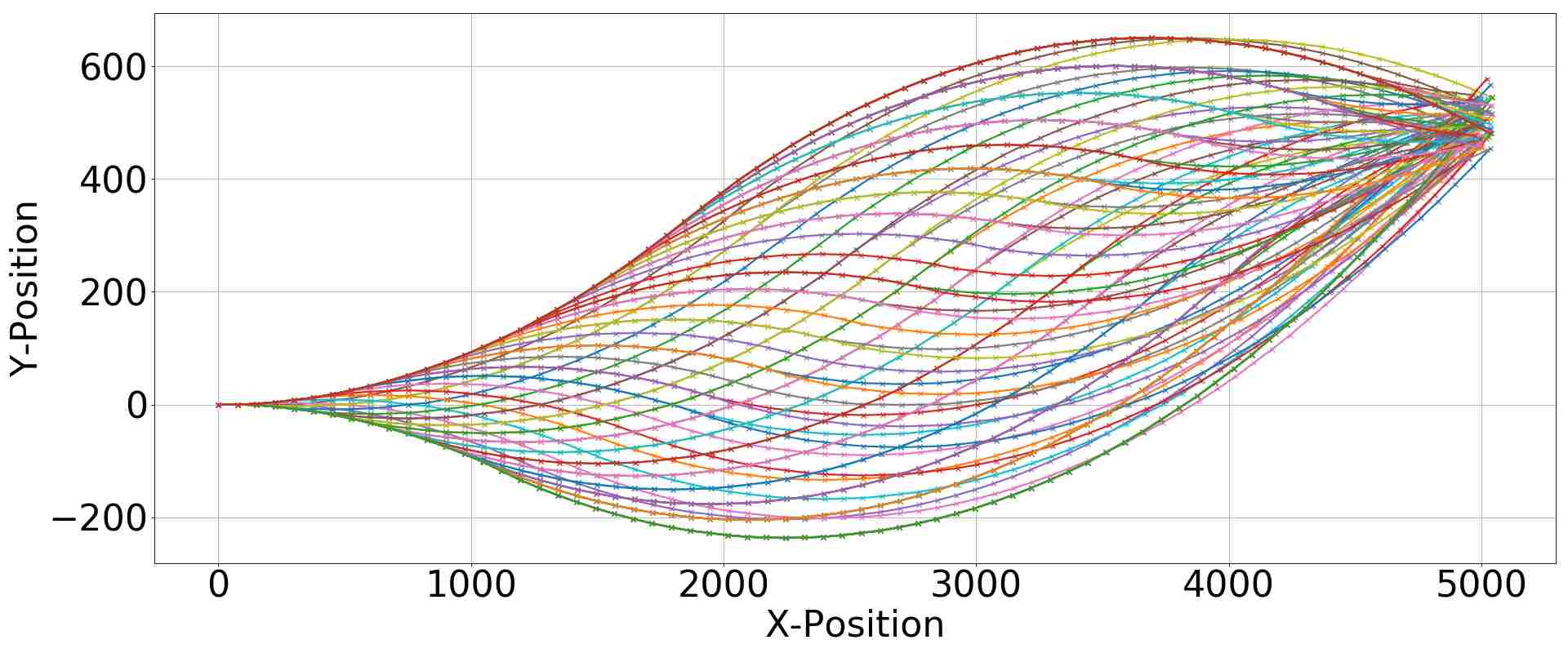}
  }\qquad
  \subfloat[Optimal motion plans for one and two rotations]{
    \label{fig:strategy-example}
    \includegraphics[width=0.45\linewidth]{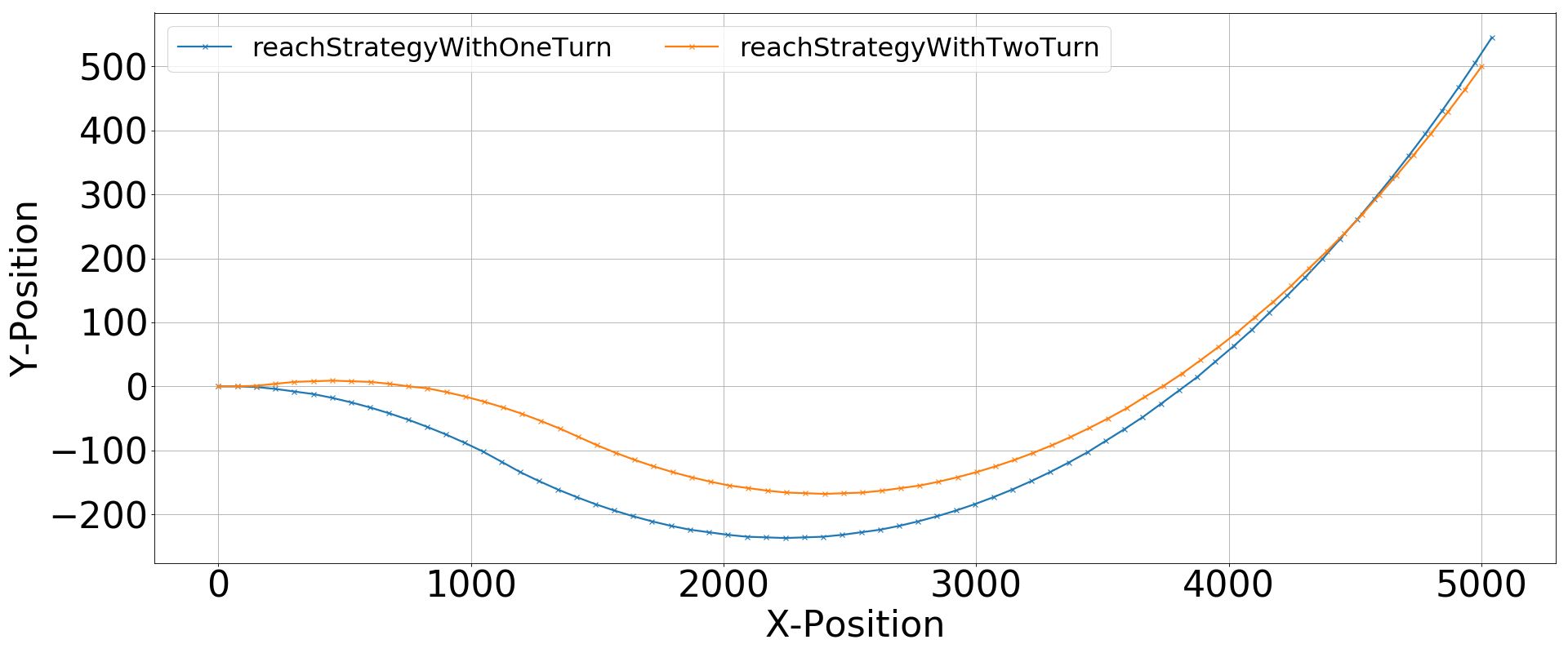}
  }
  \caption{Example motion plans derived from a synthesized strategy}
	\label{fig:strategy-examples}
\end{figure}

In illustration, Figure \ref{fig:entire-strategy-example} shows an example of the entire winning strategy --- the x and y coordinates of all possible motion plans --- to reach a target point with a permitted deviation of $100$; this target is not reachable with zero rotations.
Figure \ref{fig:strategy-example} displays the two motion plans of Figure \ref{fig:entire-strategy-example} for reaching the target with the smallest possible deviation for one and two rotations.
As both figures show, depending on the adjustment of the bevel-tip, the needle moves either in positive or in negative y direction on a circular path through the tissue.

\begin{figure}[t]%
 \centering
 \includegraphics[width=0.8\linewidth]{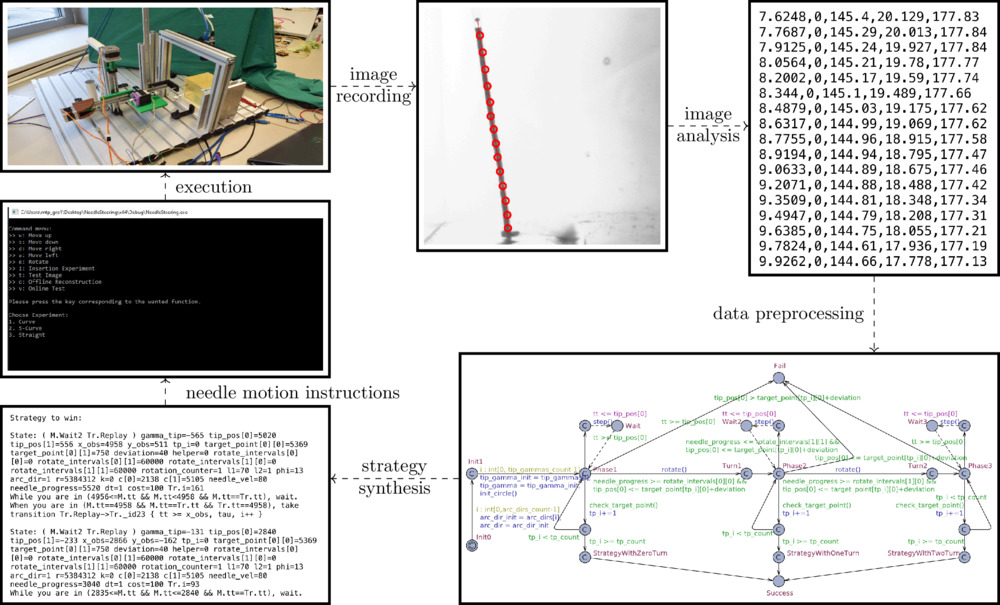}
 \caption{The workflow of strategy synthesis for needle steering in gelatin phantoms}
 \label{fig:workflow}
\end{figure}


\subsection{Using Strategies in a Real System}
Applying the generated strategies in the real experiments is not directly possible.
The biggest challenge comes from the experimental setup: since the needle is inserted manually, the actual insertion angle cannot be directly represented in the model.
Without a precise insertion angle, however, the computed motion path inevitably diverges.
We could arrange in our model for all possible insertion angles; yet, that results in a huge state space.
Instead, we separate our original model into two models, an initializing timed game model (Model 1) and a strategy timed game model (Model 2).

The sole purpose of the initial timed game is to determine the initial insertion angle. This initial game is based on a model that does not allow any rotation, but multiple insertion angles and directions.
Assuming the needle tip is already inserted and initial observed position data exists, Model 1 is initialized with the data points as targets, and the winning condition is to reach the targets including the current needle tip position as close as possible.
For this, the exact injection angle needs be determined, which is the output of the strategy.
In contrast, Model 2 --- which serves our actual strategy synthesis --- allows up to two rotations, while the insertion angle and direction are fixed via Model 1.

For the initialization, we proceed in two steps.
First, we insert the needle about $5mm$ into the tissue and record images of the needle, preprocess the images externally to extract the needle tip positions, and convert the position data into the integer domain, as required by Uppaal.
From the position data, then, we calculate the actual insertion angle of the needle using the initial timed game of Model 1.
Once we have determined the insertion angle, we initialize the strategy model with that angle and the actual needle tip position.
Figure~\ref{fig:workflow} depicts the entire workflow of our strategy synthesis for needle steering in gelatin phantoms.

\section{Experiments} \label{sec:experiments}

\begin{figure}[t]%
  \centering
  \subfloat[Strategy with one rotation ($d = 50, n_{p}=2$)]{
    \label{fig:strategies-one-rot}
    \includegraphics[width=0.45\linewidth]{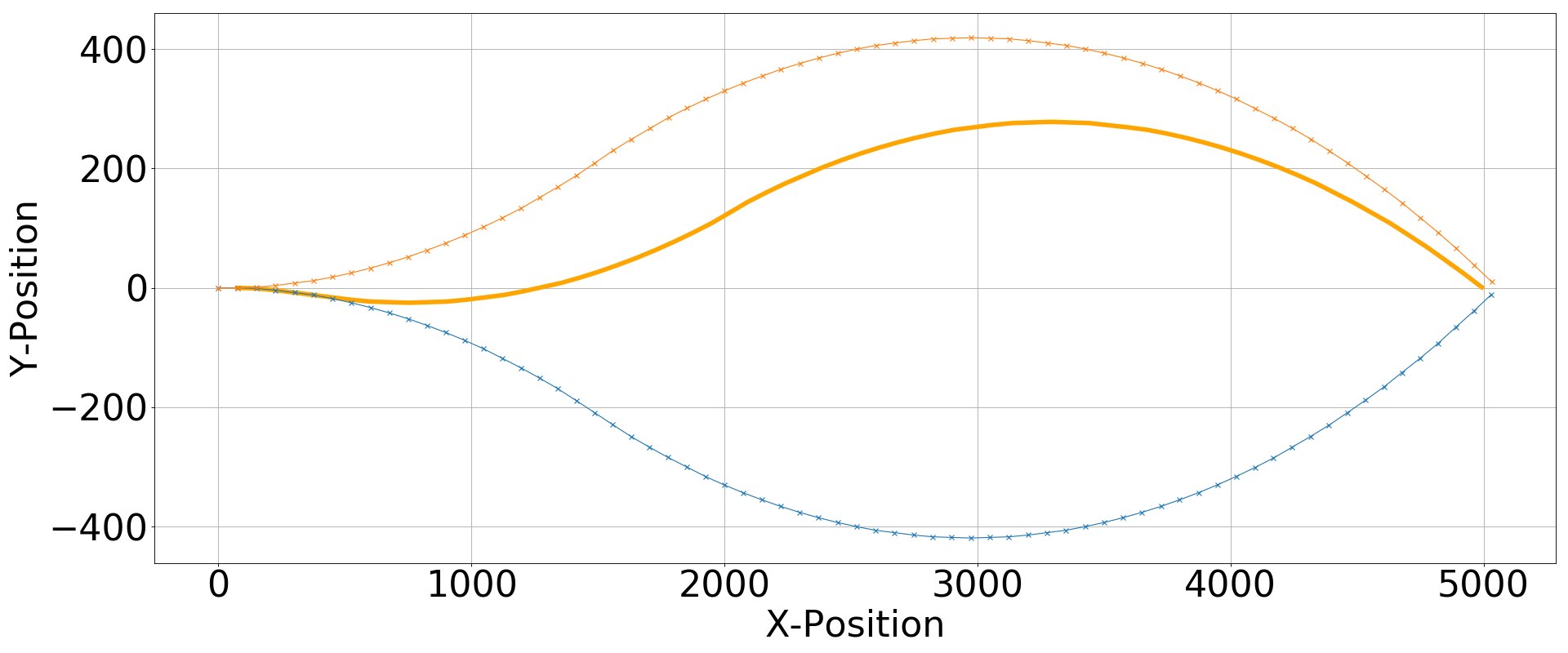}
  }\qquad
  \subfloat[Strategy with two rotations ($d = 50, n_{p}=2$)]{
    \label{fig:strategies-two-rot}
    \includegraphics[width=0.45\linewidth]{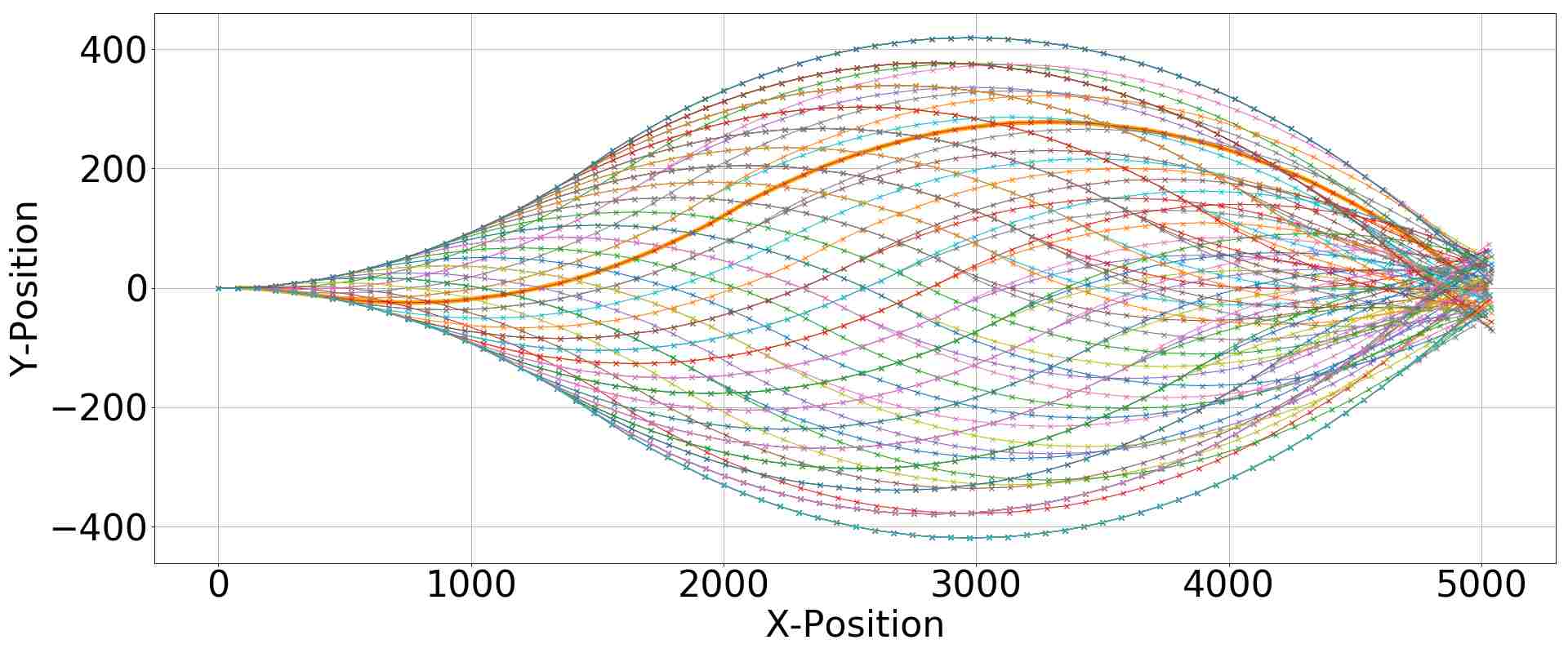}
  }\\
  \subfloat[Strategy with decreased deviation ($d = 20, n_{p}=2$)]{
    \label{fig:strategies-low-dev}
    \includegraphics[width=0.45\linewidth]{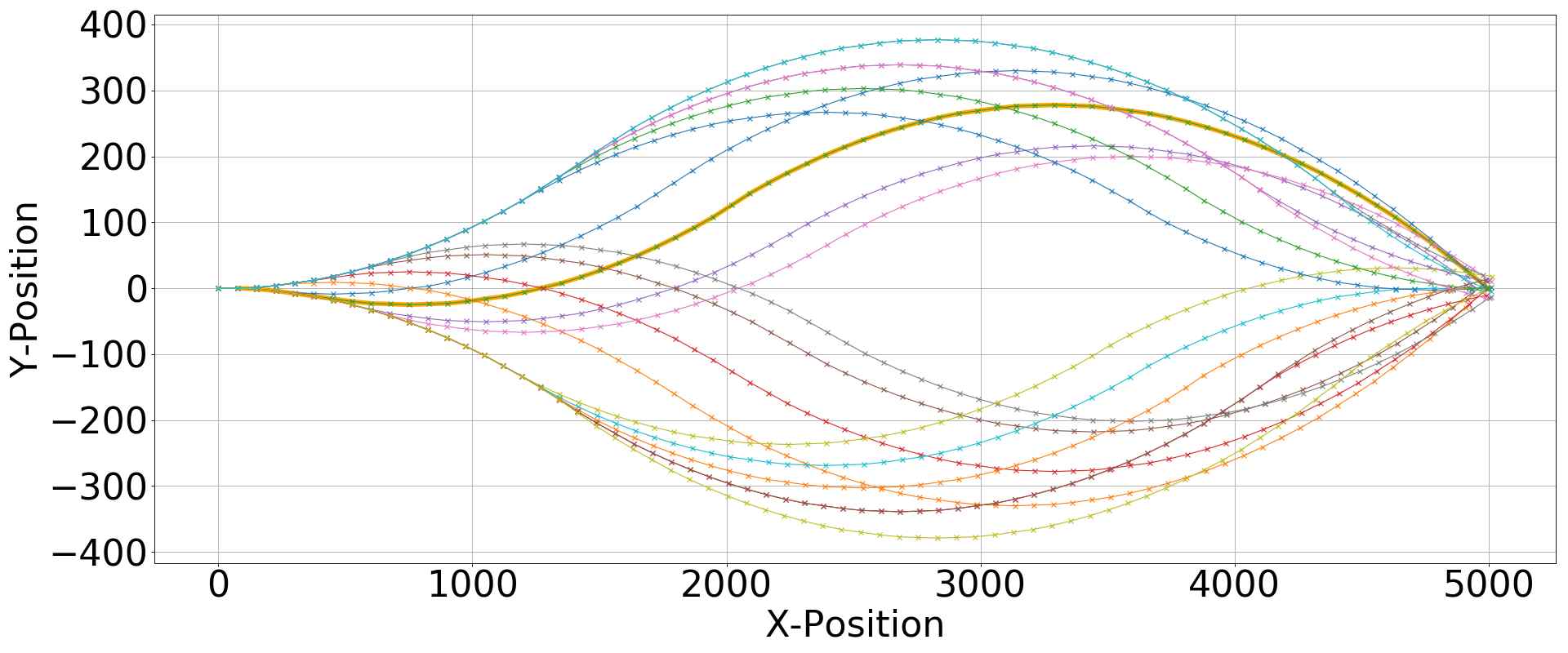}
  }\qquad
  \subfloat[Strategy with multiple target points ($d = 50, n_{p}=5$)]{
    \label{fig:strategies-multi-point}
    \includegraphics[width=0.45\linewidth]{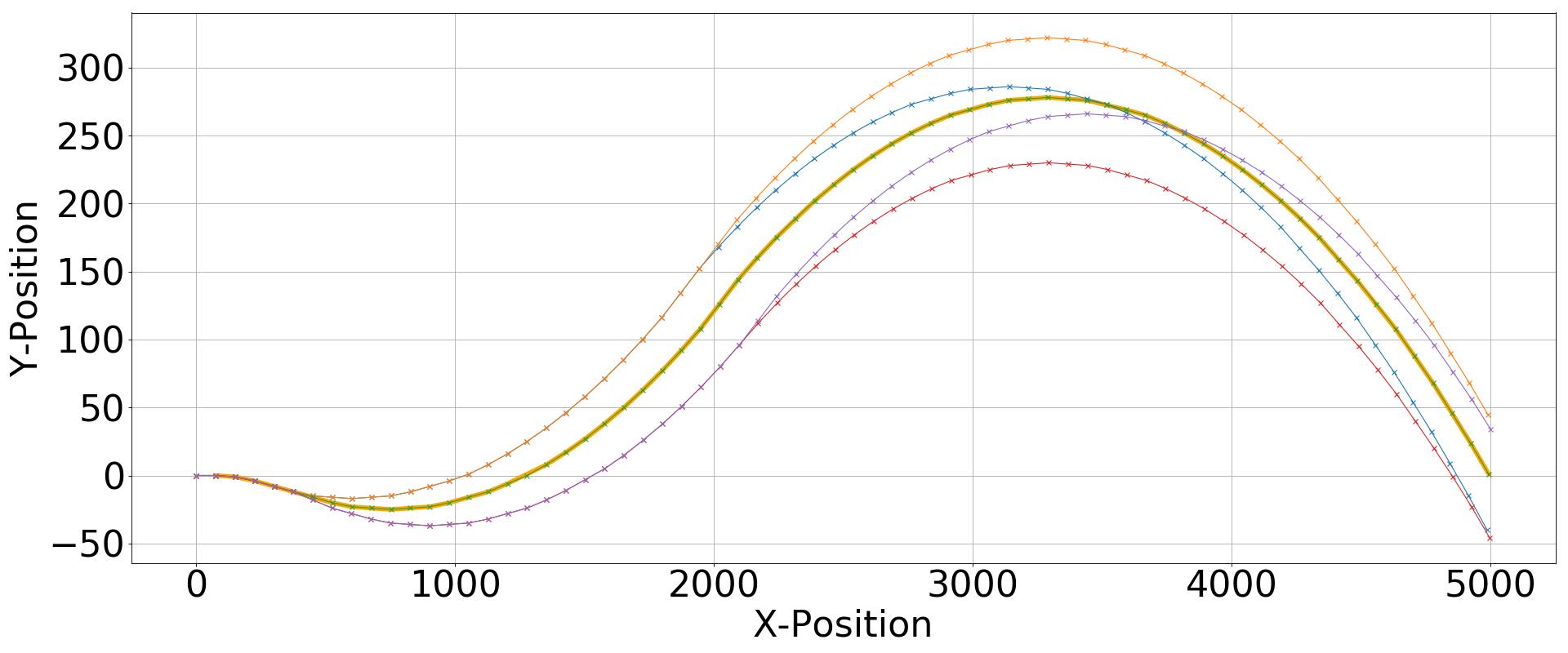}
  }\\
	\caption{Synthesized strategies for a target trace (orange) for different deviations and target point counts}
	\label{fig:strategy-plots}
\end{figure}

Based on the developed needle steering model and described strategy synthesis methods of Uppaal Stratego, we conduct two types of experiments:
First, we investigate the influence of different model parameters, i.e., the allowed number of rotations, the number of (equally distributed) target points along the needle path, and the size of the allowed deviation region around the target points, on the quality and number of generated motion plans.
Second, we apply strategy synthesis to a number of data traces gathered using the medical hardware described in Sec.~\ref{sec:medical-background}, validate that we can synthesize at least one motion plan that matches the actions performed during the individual experiments, and analyze the precision of the generated traces compared to the observed ones.

\subsection{Experiment 1: Influence of Model Parameters on Strategy Synthesis}
In the first experiment, we determine the impact of concrete model parameters on the strategy synthesis results.
To exclude the data uncertainties that necessarily come along with real measured data, the strategy synthesis for this experiment is based on data traces that are generated by simulation runs of the model itself.
The workflow of the experiment consists of the following steps:
\begin{enumerate}
  \setlength\itemsep{-0.1em}
  \item Execute a single arbitrary simulation run of the model to obtain a target trace $t_{tgt}$
  \item Select $n_{p}$ target data points of $t_{tgt}$, the number $n_{r}$ of demanded rotations, and the allowed deviation $d$ from each target point
  \item Generate the possible motion plans leading through all target data points
\end{enumerate}
Following the first step, one such trace generated by the model is shown in the plots of Figure~\ref{fig:strategy-plots}.
Using the final point of this specific trace as our target with an allowed deviation of $50$, Figure~\ref{fig:strategies-one-rot}, and Figure~\ref{fig:strategies-two-rot} show the traces of all possible synthesized motion plans for the case of one and two required needle rotations during the insertion process.
This configuration results in the total number of $2$ and $101$ motion plans, respectively.
In this example the insertion point as well as the target point lie on the x axis, which results in a symmetrical distribution of motion plans.
Note, that a different choice of target points can result in a non-symmetrical distribution.
Figure~\ref{fig:strategies-low-dev} shows the generated $20$ motion plans for a decreased deviation of $20$, and Figure~\ref{fig:strategies-multi-point} depicts those $5$ motion plans that lead through additional required data points.
We observe that the reduction of the number of needle rotations and the deviation value, as well as the increase of targeted data points all lead to a reduced amount of possible motion plans.
The data of Table~\ref{tab:experiment-1-result-table} indicates that a further adaption of these parameters finally leads to the case in which the only motion plan left is the originally generated trace.
We obtain at least one strategy for each parameterization due to the fact that our target trace is directly derived from the model, and we will indeed notice during the second experiment that the use of real data may as well lead to cases without any identifiable strategy.

\begin{table}[b]%
  \centering
	\includegraphics[width=0.65\linewidth]{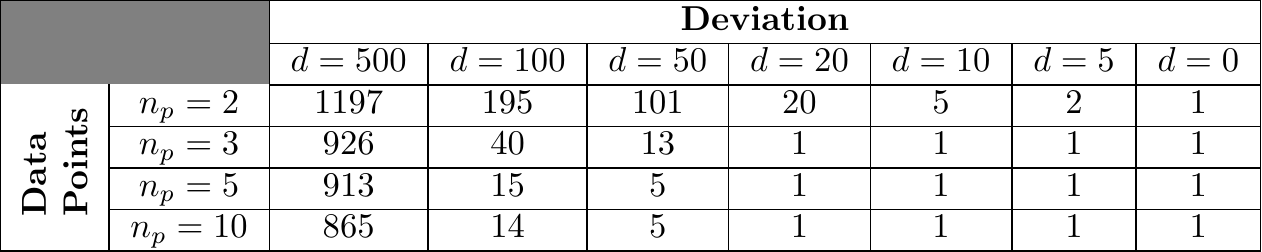}
	\caption{The number of synthesized motion plans for different deviations and numbers of data points}
	\label{tab:experiment-1-result-table}
\end{table}

\begin{table}[b]%
  \centering
	\includegraphics[width=0.85\linewidth]{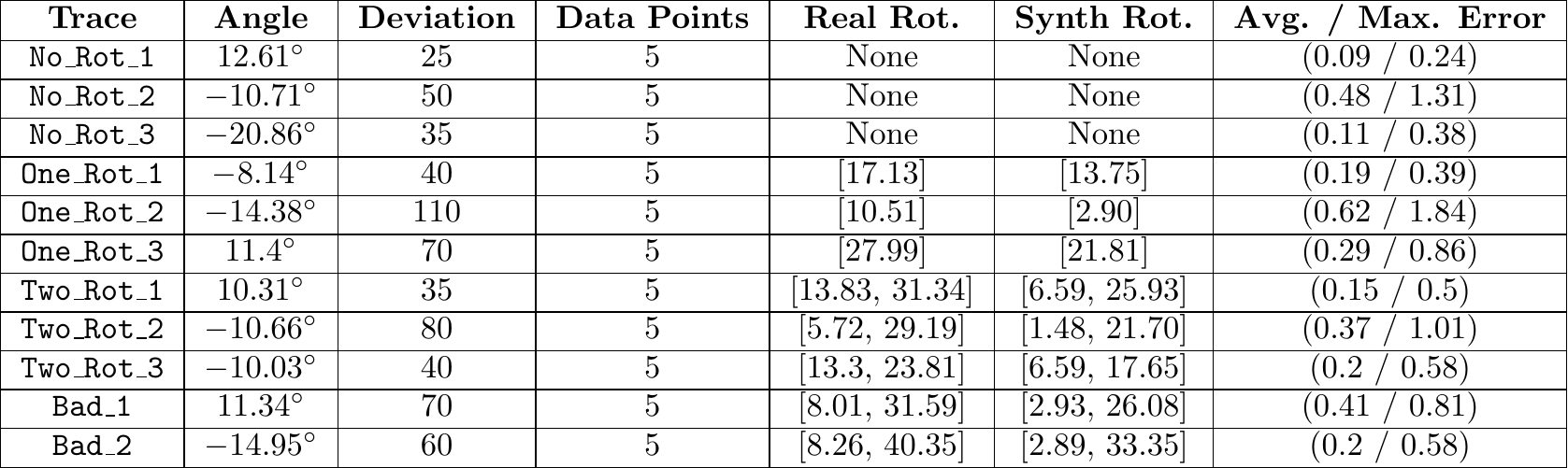}
	\caption{The strategy data for different observation data traces}
	\label{tab:experiment-2-result-table}
\end{table}

\begin{figure}[t]%
  \centering
  \subfloat[A selection of observation traces]{
    \label{fig:observation-traces}
    \includegraphics[width=0.45\linewidth]{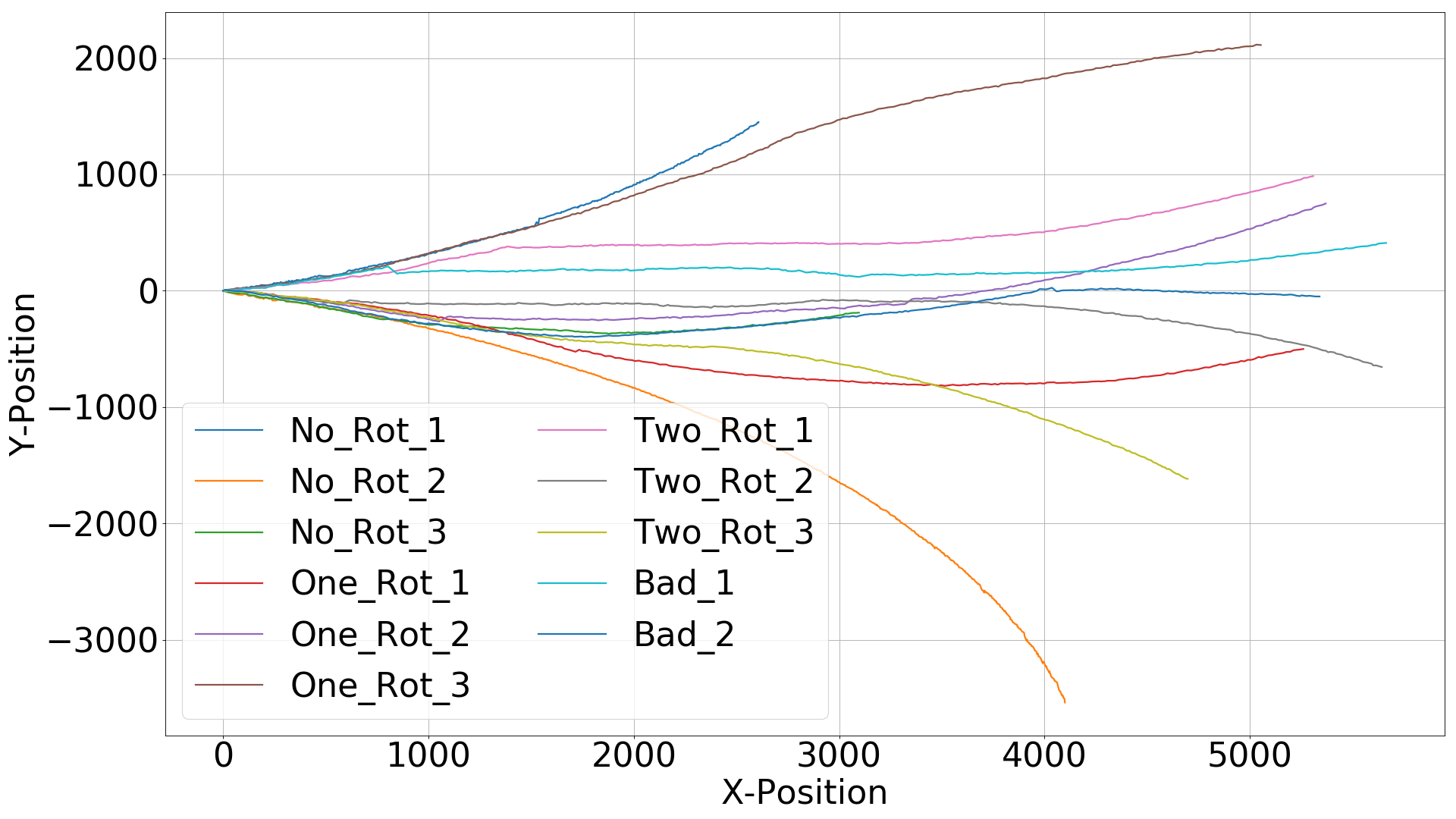}
  }\qquad
  \subfloat[The optimal strategies for the four observation traces \texttt{No\_Rot\_1}, \texttt{One\_Rot\_1}, \texttt{Two\_Rot\_1}, and \texttt{Bad\_1}]{
    \label{fig:optimal-strategies}
    \includegraphics[width=0.45\linewidth]{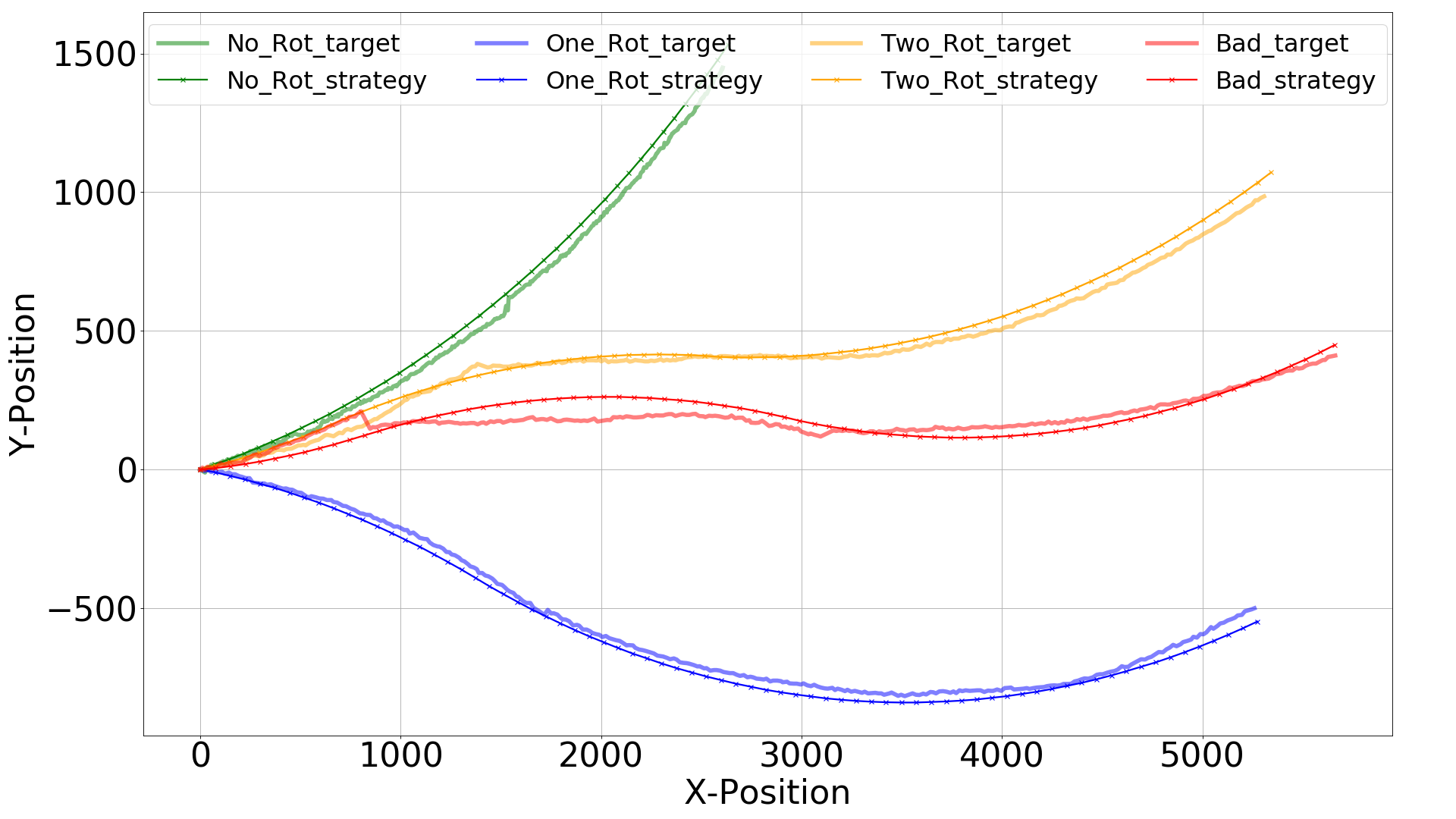}
  }
  \caption{Observation trace data and synthesized motion plans}
	\label{fig:observation-data-analysis}
\end{figure}

\subsection{Experiment 2: Offline Strategy Synthesis for Experiment Traces}
While the previous experiment was solely based on synthesized data generated by the model, we now perform a second experiment based on real measured data obtained via the needle insertion device described in Sec.~\ref{sec:medical-background}.
We will demonstrate for a set of observation data traces that a strategy can be synthesized for each trace which fits the data, and that the determined traces deviate from the observations by not more than $1.84mm$, which is a comparable accuracy for gelatin-based needle insertions \cite{Shahriari:2017}.
For an impression of the underlying data, a plot of all analyzed observation traces is shown in Figure~\ref{fig:observation-traces}, which contains traces for single and double rotations as well as both precise and erroneous measurements.
Based on these data sets, we perform the following steps:
\begin{enumerate}
  \setlength\itemsep{-0.1em}
  \item Adapt the data of a measured needle trace and add it to the Uppaal model
  \item Choose suitable parameter values described in Exp. 1, where the number of required rotations can be obtained from the trace data directly, and the initial insertion angle can be obtained via Model 1
  \item Generate the strategy that match the observation trace within given deviation bounds (if multiple or no motion plans exist, adjust and repeat Step 2)
  \item Derive the rotation points from the selected strategy
  \item Compare the strategy actions with the actually performed actions during real data acquisition
\end{enumerate}
The results of all experiment runs are gathered in Table~\ref{tab:experiment-2-result-table}, which provides the determined initial angle, allowed deviation and number of data points, the actual and identified rotation points, as well as the average and maximum error between the strategy and the observation.
All strategies describe needle paths with a maximum deviation of $1.84mm$ from the observations for the given data traces.
We observe that the synthesized rotation points compared to the real ones are systematically shifted to an earlier position.
For a reduction of the output to a single strategy, deviation values between $25$ and $110$ were used around $5$ target points.
Based on the table data, Fig~\ref{fig:optimal-strategies} exemplarily shows the best fitting strategies for the observation traces \texttt{No\_Rot\_1} (no rotation), \texttt{One\_Rot\_1} (one rotation), \texttt{Two\_Rot\_1} (two rotations), and \texttt{Bad\_1} (erroneous).
We can see that the suggested rotation points and paths fit the observation, which - in a future online application - will allow us to determine navigation strategies to arbitrary target points.

\subsection{Threats to Validity}

Potential threats to the validity of our results can be grouped into experimental issues and modeling issues. Experiments with gelatin, to begin with, always suffer from the fact that its texture makes precise reproducibility impossible.
In our specific experiments, reproducibility is additionally hampered since the initial needle insertion needs to be done manually, which invariably leads to variations in the injection angle.
Next, the quality of the image data of the needle motion is important.
Two cameras track the needle by the steel ball marker at the tip.
Yet, the steel ball marker is not accurately fixed at the tip, which results in inaccurate image recording of the needle movement, especially at the points of rotation.
Furthermore, images from an experiment often cannot be interpreted at all because there are cracks in gelatin;
we compensated for the otherwise reduced data basis by running additional experiments.
Even when sequences of useful images are available, image resolution might be insufficient and some images might simply be missing.
We mitigated that problem by checking manually for completeness. Lastly, coordinates are given as floating-point numbers and can be subject to rounding errors.
For the experiments themselves, a possible validity threat comes from the fact that we fixed the speed with which the needle is steered.
As long as we assume that tissue does not deflect a needle, however, we expect that the needle velocity has little impact only.

Another set of validity issues is due to our modeling assumptions.
We assume two dimensions, homogeneous soft tissue without anatomic obstacles, and one tissue layer, while the reality provides for three dimensions, inhomogeneous tissues, and multiple layers of tissue.
We plan to extend our models correspondingly and in fact expect the results to change for those extended models.
We also assume not more than two rotations, which might not apply in the clinical practice; yet, here we argue that low numbers of rotations certainly are desirable from a clinical perspective.
Finally, the used discretization step size might be too fine- or coarse-grained for other observation data, experiments, and applications.

\section{Conclusion and Future Work} \label{sec:conclusion}

In this paper we present a timed game model of the needle steering process in gelatin phantom using a non-holonomic model.
Based on this model we are able to synthesize strategies from which we derive motion plans to steer the needle to specific target points in gelatin.
We show that preprocessing is required to apply strategy synthesis in the real experiment and outline why an initialization step of the model---which determines the needle injection angle---is needed for the successful synthesis of strategies.

The experimental results shown that, depending on the choice of model parameters, the quality and number of generated motion plans varies and that for all sample observations at least one motion plan exists fitting the observation trace sufficiently accurate with a maximum deviation of $1.84mm$.
The synthesized rotation points, however, do not correspond exactly to the actual rotations; we conjecture inaccuracies in the experimental setting or in our modeling assumptions to be reason.

In future work we want to adapt our strategy synthesis methods so that they can be embedded in an online feedback cycle where strategies are generated at run time of the application and for bounded time intervals.
Furthermore, we plan to work on a 3-dimensional model of needle steering and a physical model of needle steering that can take anatomic obstacles and tissue inhomogeneities into account.
Lastly, we will investigate optimization problems so that strategies are generated that minimize, for example, navigation time, path length, or the number of rotations.

\bibliographystyle{eptcs}
\bibliography{bib/references.bib}

\end{document}